\documentclass[a4paper,11pt]{article}
\usepackage{latexsym,amssymb,amsfonts,amsthm,amscd,amsmath,multirow}
\usepackage[]{graphicx}
\usepackage{lineno}
\usepackage[super,numbers]{natbib}
\usepackage[symbol]{footmisc}
\usepackage[hidelinks]{hyperref}

\title{Mineral dust increases the habitability of terrestrial planets
  but confounds biomarker detection}
\author{Ian A. Boutle$^{a,b}\footnote{e-mail: ian.boutle@metoffice.gov.uk}$,
 Manoj Joshi$^c$, F. Hugo Lambert$^a$, Nathan J. Mayne$^a$,\\
 Duncan Lyster$^a$, James Manners$^{a,b}$, Robert Ridgway$^a$,
 Krisztian Kohary$^a$\\
 $^{a}${\small College of Engineering, Mathematics and Physical Sciences, University of Exeter, Exeter, EX4~4QL, UK}\\
 $^{b}${\small Met Office, FitzRoy Road, Exeter, EX1~3PB, UK}\\
 $^{c}${\small School of Environmental Sciences, University of East Anglia, Norwich, NR4~7TJ, UK}}
\date{}
\begin{document}

\maketitle

\section*{Abstract}

Identification of habitable planets beyond our solar system is a key
goal of current and future space missions. Yet habitability depends
not only on the stellar irradiance, but equally on constituent parts
of the planetary atmosphere. Here we show, for the first time, that
radiatively active mineral dust will have a significant impact on the
habitability of Earth-like exoplanets. On tidally-locked planets, dust
cools the day-side and warms the night-side, significantly widening
the habitable zone. Independent of orbital configuration, we suggest
that airborne dust can postpone planetary water loss at the inner edge
of the habitable zone, through a feedback involving decreasing ocean
coverage and increased dust loading. The inclusion of dust
significantly obscures key biomarker gases (e.g.~ozone, methane) in
simulated transmission spectra, implying an important influence on the
interpretation of observations. We demonstrate that future
observational and theoretical studies of terrestrial exoplanets must
consider the effect of dust.

\section*{Introduction}

Even before the discovery of the first potentially-habitable
terrestrial exoplanets \citep{BorAF13}, researchers have speculated on
the uniqueness of life on Earth. Of particular interest are tidally
locked planets, where the same side of the planet always faces the
star, since this is considered the most likely configuration for
habitable planets orbiting M-dwarf stars\cite{KasWR93,Bar17}, which
make up the majority of stars in our galaxy. In the absence of
observational constraints, numerical models adapted from those
designed to simulate our own planet have been the primary tool to
understand these extra-terrestrial worlds
\cite{JosHR97,TurLS16,BouMD17,DelWA19,FauTW20}. But most studies so
far have focussed on oceanic aquaplanet scenarios, because water-rich
planets are one of the likely outcomes of planetary formation
models\cite{TiaS15}, the hydrological cycle is of key importance in
planetary climate, and the definition of habitability requires stable
surface liquid water.

For a planet's climate to be stable enough for a sufficiently long
period of time to allow the development of complex organisms
(e.g. around 3 billion years for Earth\cite{NisS01}), the presence of
significant land cover may be required. The carbon-silicate weathering
cycle, responsible on Earth for the long-term stabilisation of CO$_2$
levels in a volcanic environment, acts far more efficiently on land
than at the ocean floor \citep{AbbCC12}. Some studies have attempted
to simulate the effects of the presence of land
\citep{Jos03,AbeAS11,YanBF14,LewLB18,WayDA18}, demonstrating how it
would affect the climate and atmospheric circulation of a
tidally-locked planet, such as Proxima~b\cite{TurLS16,DelWA19}. More
specific treatments of land surface features such as topography have
only been briefly explored \cite{DelWA19,WayDK16}.

Mineral dust is a significant component of the climate system whose
effects have been hitherto neglected in climate modelling of
exoplanets. Mineral dust is class of atmospheric aerosol lifted from
the planetary surface and comprising the carbon-silicate material
which forms the planetary surface (it should not be conflated with
other potential material suspended in a planetary atmosphere, such as
condensable species (clouds) or photochemical haze). Dust is raised
from any land surface that is relatively dry, and free from
vegetation. Dust can cool the surface by scattering stellar radiation,
but also warm the climate system through absorbing and emitting
infra-red radiation. Within our own solar system, dust is thought to
be widespread in the atmosphere of Venus \citep{GreAE92}, and is known
to be an extremely important component of the climate of Mars, which
experiences planetary-scale dust storms lasting for weeks at a time
\citep{Zur82,KahMH06}. Even on Earth, dust can play a significant role
in regional climate \citep{LamDP08,KokWM18} and potentially in global
long-term climate \citep{RidW02}.

Here we demonstrate the importance of mineral dust on a planet's
habitability. Given our observations of the solar system, it is
reasonable to assume that any planet with a significant amount of dry,
ice- and vegetation-free land cover, is likely to have significant
quantities of airborne dust. Here we show for the first time that
mineral dust plays a significant role in climate and habitability,
even on planets with relatively low land fraction, and especially on
tidally-locked planets. We also show that airborne dust affects near
infra-red transmission spectra of exoplanets, and could confound
future detection of key biomarker gases such as ozone and
methane. Airborne mineral dust must therefore be considered when
studying terrestrial exoplanets.

\section*{Results}
\subsection*{Schematic mechanisms}

We consider two template planets, a tidally-locked planet orbiting an
M-dwarf (denoted TL), with orbital and planetary parameters taken from
Proxima~b, and a non-tidally-locked planet orbiting a G-dwarf (denoted
nTL), with orbital and planetary parameters taken from Earth. The
choice of parameters is merely to give relatable examples, the results
presented are generic and applicable to any planet in a similar
state. We also consider the planets to be Earth-like in atmospheric
composition, i.e. 1~bar surface pressure and a nitrogen dominated
atmosphere, as this is the most well understood planetary atmosphere,
and only one known to be inhabited. For each of these planets we
consider a range of surface land-cover amounts and configurations,
designed to both explore the parameter space that may exist and
understand in which scenarios dust is important. Starting from
well-understood aquaplanet simulations \citep{BouMD17} derived using a
state-of-the-art climate model \citep{WalBB19}, we increase the
fraction of land in each model grid-cell equally, until the surface is
completely land. This experiment (denoted Tiled) acts to both increase
the amount of land available for dust uplift, whilst reducing the
availability of water and thus the strength of the hydrological cycle,
without requiring knowledge of continent placement. For the TL case,
we additionally conduct simulations in which a continent of increasing
size is placed at the sub-stellar point (denoted Continents). This
produces a fundamentally different heating structure from the central
star \citep{LewLB18} and significantly increases the effect of the
dust for small land fractions, whilst allowing a strong hydrological
cycle to persist.

For each planet and climate configuration, we run two simulations, one
without dust, called NoDust, equivalent to all previous studies of
rocky exoplanets, and one in which dust can be lifted from the land
surface, transported throughout the atmosphere and interact with the
stellar and infra-rad radiation and atmospheric water, called Dust.

\begin{figure}[tbhp]
\centering
\includegraphics[width=\columnwidth]{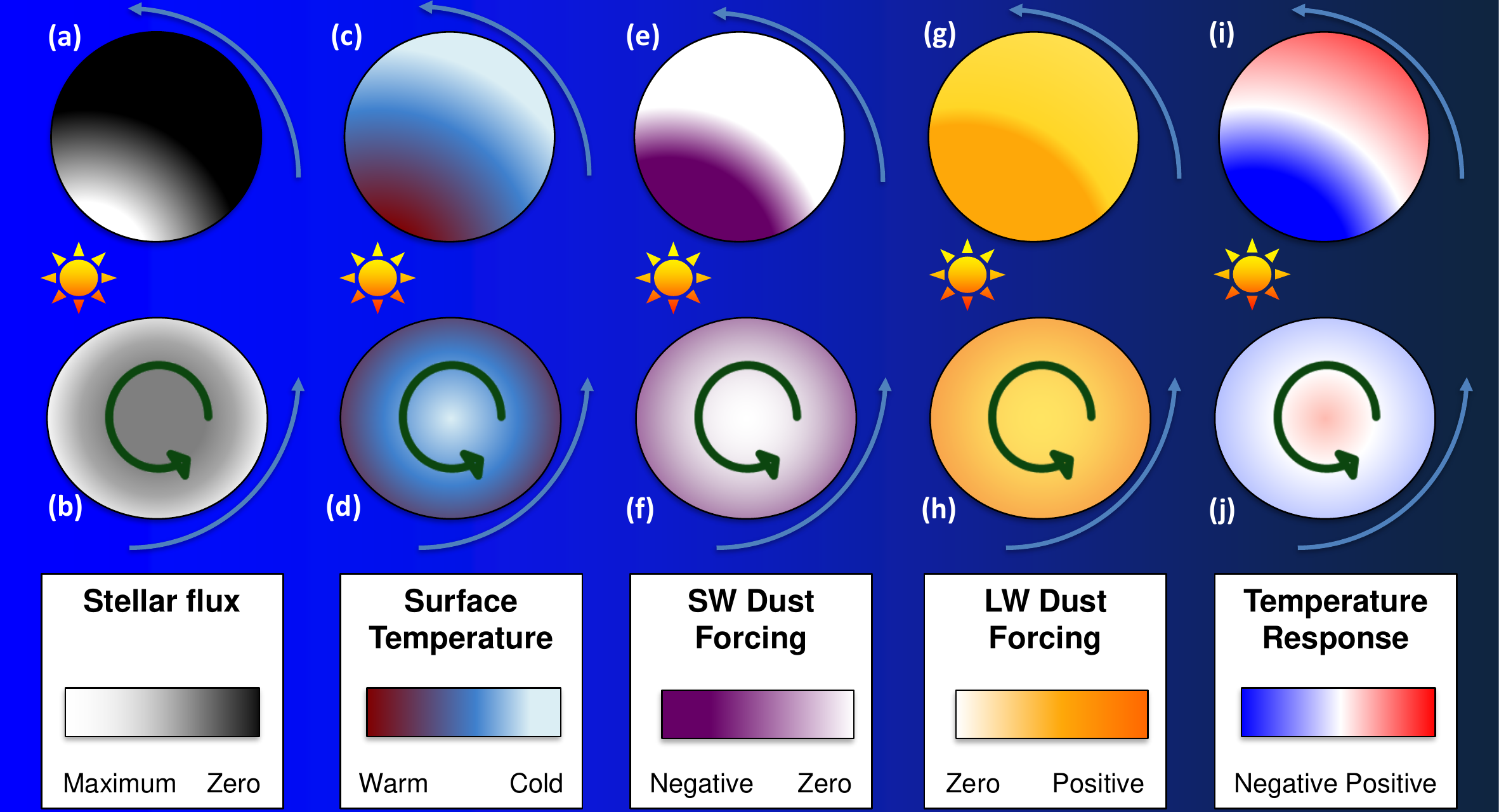}
\caption{{\bf Schematic showing the effect dust has on the climate of
    planets}. For a tidally-locked planet (a) and non-tidally-locked
  planet (b), panels a-d show the base state of the planets, e-h show
  the short-wave (stellar) and long-wave (infra-red) forcing (change
  in surface energy balance) introduced by dust, and i-j show the
  resultant effect of the forcing on the surface temperature. Blue
  arrows show the motion of the planet around the star, and green
  arrows show the rotation of the planet relative to the star.}
\label{fig-schematic}
\end{figure}

The mechanisms through which dust affects planetary climate are
illustrated in Figure~\ref{fig-schematic}. Incoming stellar radiation
is concentrated over a smaller area on the TL planet
(Fig.~\ref{fig-schematic}a) compared to the nTL case
(Fig.~\ref{fig-schematic}b). Strong surface winds on the day-side of
TL allow for much greater uplift of dust than the equatorial doldrums
of nTL. The super-rotating jet on TL is more efficient at transporting
this dust to cooler regions on the night-side
(Fig.~\ref{fig-schematic}c), than the more complex atmospheric
circulation on nTL is at transporting dust to the poles
(Fig.~\ref{fig-schematic}d). The radiative forcing, or change in
surface energy balance caused by airborne dust, is therefore weaker
for nTL than TL. As a result, the nTL planet is broadly cooled by dust
(Fig.~\ref{fig-schematic}j) because the airborne dust's infra-red
greenhouse effect (Fig.~\ref{fig-schematic}h) is cancelled out by the
stellar radiation changes due to scattering and absorption by airborne
dust (Fig.~\ref{fig-schematic}f). However, the TL planet is strongly
cooled on its warm day-side by similar mechanisms, but warmed on its
night-side (Fig.~\ref{fig-schematic}i) because the airborne dust's
infra-red greenhouse effect (Fig.~\ref{fig-schematic}g) has no stellar
radiation change to offset it (Fig.~\ref{fig-schematic}e).

\subsection*{Habitable zone changes}

\begin{figure}[tbhp]
\centering
\includegraphics[width=\columnwidth]{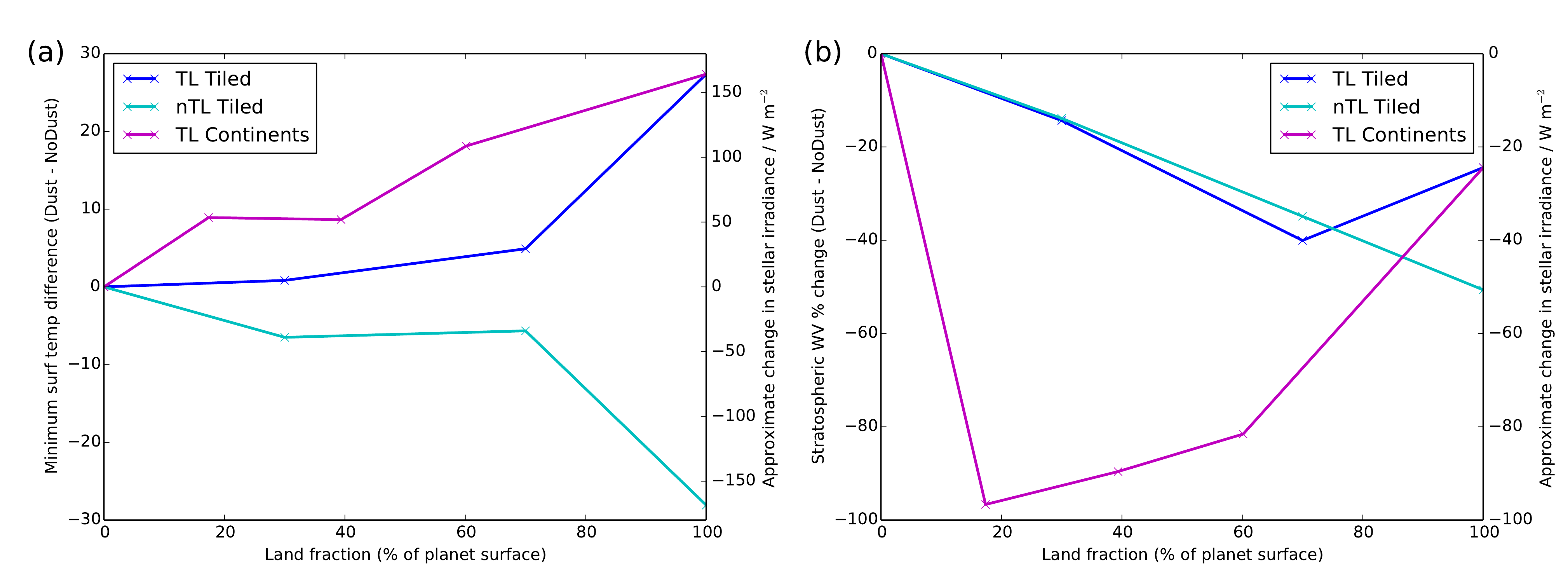}
\caption{{\bf Effect of dust on habitable zone boundary indicators.}
  Differences in (a) minimum surface temperature, and (b)
  stratospheric ($\approx 50$~hPa) water-vapour content, between
  simulations with (Dust) and without (NoDust) mineral dust, as a
  function of land fraction. The different planetary and surface
  setups are shown in the legend. Approximate equivalent changes in
  stellar irradiance required to achieve similar responses in a dust
  free planet are shown on the right axis.}
\label{fig-land_frac}
\end{figure}
Figure~\ref{fig-land_frac} shows two key metrics we use to quantify
the outer and inner edges of the habitable zone for our template
planets. The outer edge of the habitable zone is likely to be
controlled by the temperature at which CO$_2$ condenses
\citep{TurFL17}, which for the concentrations and surface pressures
considered here, is at $\approx 125$~K. Keeping the minimum
temperature above this threshold is therefore a key requirement to
maintaining a CO$_2$ greenhouse effect, and preventing a planet's
remaining atmospheric constituents from condensing
out. Figure~\ref{fig-land_frac}a shows that for the TL case, the
presence of dust always acts to increase the minimum temperature found
on the planet (blue and magenta lines). The effect of dust is to
sustain a greenhouse effect at a lower stellar irradiance than when
dust is absent, implying that dust moves the outer edge of the
habitable zone away from a parent star. The effect is not especially
sensitive to the specific arrangement of the land (magenta vs blue
lines in Fig.~\ref{fig-land_frac}a), but is very sensitive to fraction
of the surface covered by land; the approximate change in stellar
radiation at the outer edge of the habitable zone is over
$150$~W~m$^{-2}$ for a totally land-covered planet, but even up to
$50$~W~m$^{-2}$ for a planet with the same land coverage as Earth.
Such results are in stark contrast to the nTL case, for which dust
always acts to reduce the minimum surface temperature (cyan line in
Fig.~\ref{fig-land_frac}a), moving the outer edge of the habitable
zone inwards.

The inner edge of the habitable zone is likely to be controlled by the
rate at which water-vapour is lost to space, often termed the moist
greenhouse \citep{Kas88,ZsoSW13,LecFC13}. The strength of the water
vapour greenhouse effect increases with surface temperature,
eventually leading to humidities in the middle atmosphere that are
large enough to allow significant loss of water to
space. Stratospheric water-vapour content is therefore a key indicator
of when an atmosphere will enter a moist
greenhouse. Figure~\ref{fig-land_frac}b shows that for all our
simulations the effect of dust is to reduce stratospheric water-vapour
content, i.e. dust suppresses the point at which a moist greenhouse
will occur and moves the inner edge of the habitable zone nearer to
the parent star. The effect on the habitable zone can be approximately
quantified by utilising additional simulations done with increased or
reduced stellar flux and a constant tiled land fraction of $70$\%
(Table~\ref{tab-expts}). They show that stratospheric water-vapour
scales approximately logarithmically with stellar flux, allowing us to
infer that the $30-60$\% reduction in stratospheric water-vapour
caused by dust (shown in Fig.~\ref{fig-land_frac}b) roughly
corresponds to a stellar flux reduction of $30-60$~W~m$^{-2}$. In
contrast to the effect on the outer edge, both our TL and nTL
simulations result in a reduction in stratospheric water-vapour when
including dust, demonstrating that the inward movement of the inner
edge of the habitable zone is a ubiquitous feature of atmospheric
dust. However, here the magnitude of the effect is more dependent on
the arrangement of the land, and therefore more
uncertain. Supplementary Notes 1 and 2 give more details on this.

In summary radiatively active atmospheric dust increases the size of
the habitable zone for our tidally-locked planets, both by moving the
inner edge inwards and outer edge outwards. For our non-tidally-locked
planets, both the inner and outer edges of the habitable zone move
inwards, so the consequences for habitable zone size depend on which
effect is stronger. The exact size of the habitable zone is a subject
of much debate \citep{KasWR93,ZsoSW13,LecFC13,KopRK13}, and how well
our results can be extrapolated to previous estimates of its size are
covered in the discussion. But to illustrate the potential importance
of dust, conservative estimates from Kasting et al. \cite{KasWR93}
suggest a stellar irradiance range of $\sim 750$~W~m$^{-2}$ from inner
to outer edge. Figure~\ref{fig-land_frac}a shows that the effect of
dust is equivalent to changing the stellar irradiance by up to
$150$~W~m$^{-2}$, thereby moving the outer edge of the habitable zone
by up to $10$\% in either direction.

\begin{figure}[tbhp]
\centering
\includegraphics[width=\columnwidth]{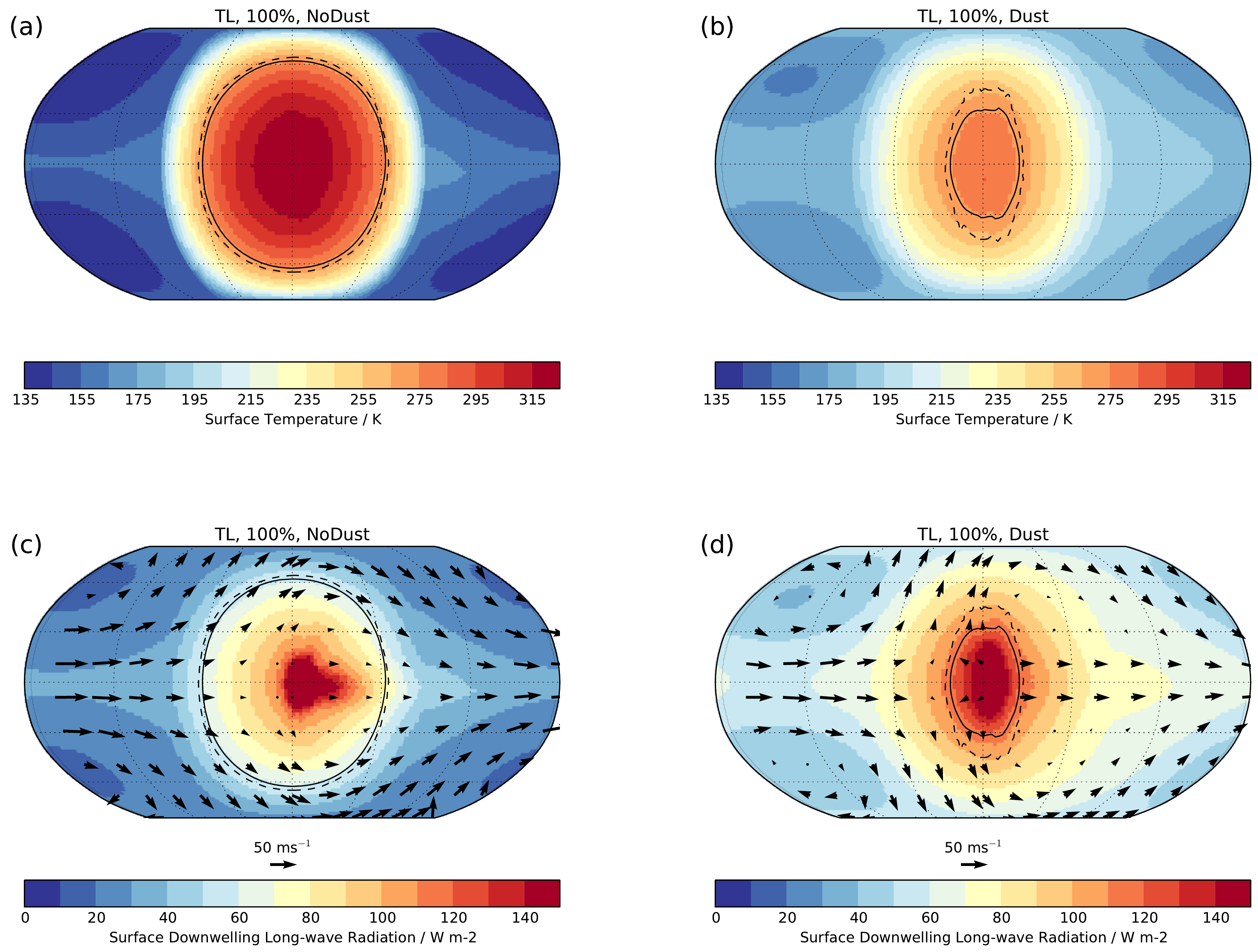}
\caption{{\bf Mechanisms driving surface temperature change.} Surface
  temperature (a-b) and surface downwelling long-wave radiation (c-d),
  from the TL case with $100$\% land cover for the NoDust (a,c) and
  Dust (b,d) simulations. Also shown are the mean (solid) and maximum
  (dashed) $273$~K contours, and wind vectors at $8.5$~km ($\approx
  300$~hPa, in c and d).}
\label{fig-TL}
\end{figure}

Figure~\ref{fig-TL} illustrates the effects of dust on climate for the
TL case in more detail. We show results for the $100$\% land
simulation where the dust effect is strongest, and although the effect
is weakened with lower fractions of land, the mechanisms remain the
same. The dust particles are lifted from the surface on the day-side
of the planet, since uplift can only occur from non-frozen
surfaces. There they are also strongly heated by incoming stellar
radiation. The larger particles cannot be transported far before
sedimentation brings them back to the surface, but the smaller
particles can be transported around the planet by the strong
super-rotating jet expected in the atmospheres of tidally-locked
planets \citep{ShoP11}. The smaller dust sizes are therefore
reasonably well-mixed throughout the atmosphere, and able to play a
major role in determining the radiative balance of the night-side of
the planet. This highlights an important uncertainty not yet discussed
-- our assumption that surface dust is uniformly distributed amongst
all size categories. As only the small- to mid-sized dust categories
play a major role in determining the planetary climate, increasing or
decreasing the amount of surface dust in these categories can increase
or decrease the quantitative effects. Similarly, the precise
formulation of the dust uplift parametrization can have a similar
quantitative effect on the results presented. More discussion is given
in Supplementary Note 1, but neither uncertainty changes the
qualitative results presented.

The coldest temperatures are found in the cold-trap vortices on the
night-side of the planet (Fig.~\ref{fig-TL}a), which without dust are
$\approx 135$~K. The effect of dust is to raise the temperature
reasonably uniformly by $\approx 25$~K across the night-side of the
planet (Fig.~\ref{fig-TL}b), significantly raising the temperature of
the cold-traps above the threshold for CO$_2$ condensation. The
increase in surface temperature arises because of a corresponding
increase in the downwelling infra-red radiation received by the
surface of the night-side of the planet (Fig.~\ref{fig-TL}c and d),
which is approximately doubled compared to a dust free case.

Supplementary Note 2 explores some of the more detailed responses of
the different land surface configurations that were shown in
Figure~\ref{fig-land_frac}. However, in all cases, the effect on the
habitable zone and mechanisms for the change are consistent with those
described above, and are likely to be significant for any continental
configuration and even for low fractions of land.

\subsection*{Simulated observations}

\begin{figure}[tbhp]
\centering
\includegraphics[width=\columnwidth]{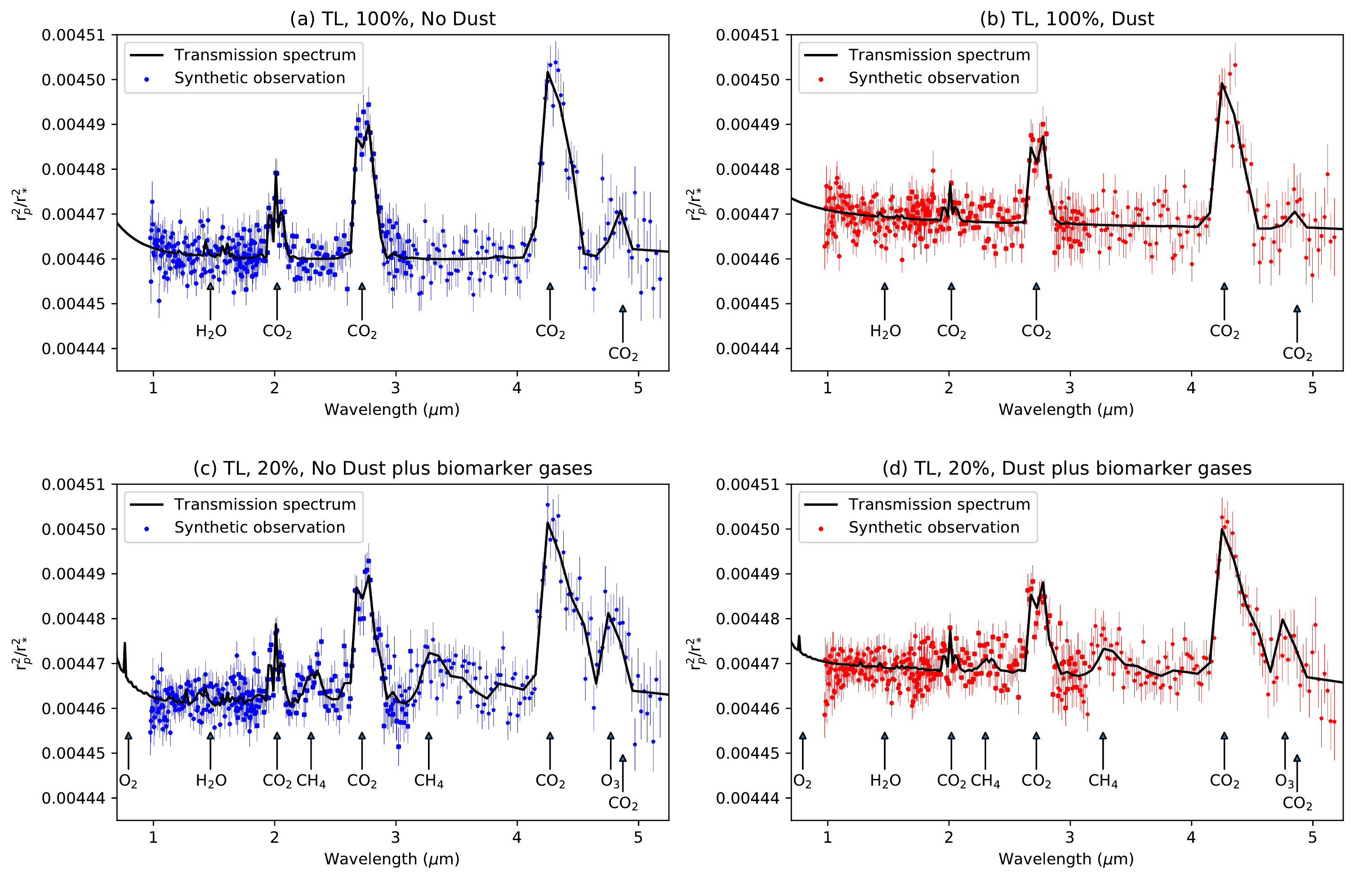}
\caption{{\bf Effect of dust on planetary observations.} Simulated
  transmission spectra (black) and synthetic JWST observations
  (blue/red), from 15 transits for a dusty (b,d) and non-dusty (a,c)
  tidally-locked planet, orbiting an M-dwarf of apparent magnitude
  similar to Proxima Centauri, with $100$\% land coverage and no
  potential biomarker gases (a-b), and $20$\% land cover and biomarker
  gases (c-d). A one standard deviation error on the synthetic
  observation is shown.}
\label{fig-jwst}
\end{figure}

A key question regarding airborne mineral dust is how it would affect
the interpretation of potential future spectra of terrestrial
exoplanets. Figure~\ref{fig-jwst} presents synthetic observations
created from our model output combined with the PandExo simulator
\citep{BatMP17} of the NIRSpec (G140M, G235M and G395M modes)
instrument on the James Webb Space Telescope (JWST), following the
method described in Lines et al. \cite{LinMM18}. We focus here on the
TL case, and compare the relatively dry $100$\% land-cover simulation
with the $20$\% land-cover arranged as a continent simulation, to
demonstrate how even planets with low dust loading and a strong
hydrological cycle can be affected. We additionally consider the
$20$\% land-cover simulation to have an atmospheric composition which
is Earth-like, i.e. it contains the key observable potential biomarker
gases oxygen, ozone and methane \citep{SchEK18} in present day Earth
concentrations. Adding these gases does not greatly affect the
climatic state \citep{BouMD17}, but can significantly alter the
observed spectra. We consider a target object with the apparent
magnitude of Proxima Centauri, as stars near this range are the most
likely candidates for observing in the near future (Proxima~b itself
does not transit \cite{JenHC19}, but that does not invalidate our
results for similar planets around similar stars). We discuss how our
results change for dimmer stars such as TRAPPIST-1 in Supplementary
Note 3.

Figure~\ref{fig-jwst} shows that airborne dust effectively introduces
a new continuum absorption into the spectrum, which completely
obscures many of the minor absorption peaks similar to previous
studies of hotter planets \citep{LinMM18,KreBD14}, some of which are
associated with potential biomarker gases, such as methane ($2.3$
and $3.3$~$\mu$m) and ozone ($4.7$~$\mu$m). An oxygen feature at
$0.76$~$\mu$m is also significantly obscured in the dusty spectrum,
and although it falls outside the spectral range of JWST, is similarly
unlikely to be prominent enough if it was within the observable
spectrum. Importantly, biomarker gas features are obscured even
when dust loading is relatively low (Fig.~\ref{fig-jwst}c,d),
i.e. even relatively wet planets with a strong hydrological cycle are
prone to having important spectral peaks being obscured from
observation by dust.

\section*{Discussion}

Given the radiative properties of dust, and the dependence of its
impact on the climate on land fraction (Figure~\ref{fig-land_frac}),
it could potentially produce a strong negative feedback for planets
undergoing significant water loss at the inner edge of the habitable
zone. As water is lost and the fraction of the surface covered by
ocean decreases, the amount of dust that is suspended in the
atmosphere will likely increase, which in turn cools surface
temperatures, quite dramatically in the case of a tidally-locked
planet, reducing the amount of water-vapour in both the lower and
middle atmosphere. Airborne dust can therefore act as a temporary
brake on water loss from planets at the inner edge of the habitable
zone in a similar manner to the ocean fraction/water-vapour feedback
\citep{AbeAS11}. However, how dust interacts with other mechanisms
affecting the inner-edge of the habitable zone requires further
study. For example, the potential bi-stable state of planets with
water locked on the night-side \cite{LecFC13b}, which may also widen
the habitable zone, may be partly offset by the presence of dust if
the warmer night-side (due to mechanisms discussed here) allows some
of the water to be liberated back to the day-side.

Estimates of the outer-edge of the habitable zone\cite{KopRK13} are
also typically made with much higher CO$_2$ partial pressures than
those considered here (up to $10$~bar). It is unclear that such high
CO$_2$ concentrations could be achieved in the presence of land, due
to increased weathering activity preventing further CO$_2$ build up
\cite{KitGM11}. If they are, the quantitative effect of dust will
depend on a range of compensating uncertainties. For example, dust
uplift should be enhanced due to higher surface stresses in a
higher-pressure atmosphere. However dust transport to the night-side
may be reduced in the weaker super-rotating jet due to reduced
day-night temperature contrasts \cite{FauTV19}.

Our results have implications for studies of the history of our own
planet before terrestrial vegetation covered large areas, with a
particular example being the faint young Sun problem of Archean Earth
\cite{ChaFW13}. The land masses which are believed to have emerged
during this period will have been unvegetated, and therefore a
significant source of dust uplift into the atmosphere if dry and not
covered in ice. As we have shown, this dust would have a cooling
effect on the planetary climate, potentially making the faint young
Sun problem harder to resolve. However, it is also possible that
microbial mats might have covered large areas of the land surface
before vegetation evolved. The exact nature of such cover, and how
much it would hinder dust lifting into the atmosphere, has yet to be
quantified.

It is clear that the possible presence of atmospheric dust must be
considered when interpreting observations. The feature-rich spectrum
observed from a dust-free atmosphere containing water-vapour, oxygen,
ozone and methane (Fig.~\ref{fig-jwst}c) is transformed into a flat,
bland spectrum where only major CO$_2$ peaks are visible above the
background dust continuum (Fig.~\ref{fig-jwst}d). Observations
returning a spectrum such as this could easily be misinterpreted as
being caused by a dry atmosphere containing only nitrogen and CO$_2$,
i.e. Fig.~\ref{fig-jwst}d interpreted as Fig.~\ref{fig-jwst}a. The
result would be a potentially very interesting planet being
characterised as dry, rocky and lifeless. On the other hand, if
spectra are obtained which can unambiguously place a limit on dust
generation, such results imply a mechanism that inhibits dust lifting,
whether it be some combination of very small land fraction,
significant ice or vegetation cover, or other dust-inhibiting
mechanism: such a result would also be of great interest to those
interpreting observations.

Finally, our results have wide-ranging consequences for future studies
of the habitability of terrestrial rocky planets. Such studies should
include models of airborne dust as well as observational
constraints. Furthermore, our results strongly support the continued
collaboration between observational and modelling communities, as they
demonstrate that observations alone cannot determine the size of the
habitable zone: it crucially depends on properties of the planetary
atmosphere, which are presently only accessible via climate modelling.


\section*{Methods}

\subsection*{Numerical model setup}

Our general circulation model of choice is the GA7 science
configuration of the Met Office Unified Model \citep{WalBB19}, a
state-of-the-art climate model which incorporates within it a mineral
dust parameterisation \citep{Woo01,Woo11}, which includes uplift from
the surface, transport by atmospheric winds, sedimentation, and
interaction with radiation, clouds and precipitation. The
parameterisation comprises 9 bins of different sized dust particles
($0.03-1000$~$\mu$m). The largest 3 categories ($>30$~$\mu$m)
represent the precursor species for atmospheric dust; these are the
large particles which are not electro-statically bound to the surface,
but can be temporarily lifted from the surface by turbulent
motions. They quickly return to the surface under gravitational
effects, and as such are not transported through the atmosphere (they
do not travel more than a few metres). However, they are important
because their subsequent impact with the surface is what releases the
smaller particles into the atmosphere. These smaller 6 categories
($<30$~$\mu$m) are transported by the model's turbulence
parameterisation \citep{LocBBM00}, moist convection scheme
\citep{GreR90} and resolved atmospheric dynamics
\citep{WooSWA14}. They can return to the surface under gravitational
settling, turbulent mixing, and washout from the convective or
large-scale precipitation schemes \citep{WilB99}. The absorption and
scattering of short- and long-wave radiation by dust particles is
based on optical properties calculated from Mie theory, assuming
spherical particles, and each size division is treated independently.

The land surface configuration is almost identical to that presented
in Lewis et al. \cite{LewLB18}, i.e. a bare-soil configuration of the
JULES land surface model set to give the planet properties of a sandy
surface. Our key difference is the use of a lower surface albedo
($0.3$). The land is at sea-level altitude with zero orography and a
roughness length of $1\times10^{-3}$~m for momentum and
$2\times10^{-5}$~m for heat and moisture (although these are reduced
when snow is present on the ground). The soil moisture is initially
set to its saturated value, but evolves freely to its equilibrium
state. Land is assumed to comprise dust of all sizes, uniformly
distributed across the range. The dust parameterisation is used in its
default Earth setup, and naturally adapts to the absence of
vegetation, suppresses uplift in wetter regions, and prevents it from
frozen or snow covered surfaces. The ocean parametrization is a slab
ocean of $2.4$~m mixed layer depth with no horizontal heat transport,
as was used in Boutle et al. \cite{BouMD17}, and includes the effect
of sea-ice on surface albedo following the parametrization described
in Lewis et al. \cite{LewLB18}. It is worth noting that whilst the
setup implies an infinite reservoir of both water in the ocean and
dust on the land, this is not actually a requirement of the results --
all that is required is enough water/dust to support that which is
suspended in the atmosphere and deposited in areas unfavourable for
uplift (e.g. the night-side of the TL planet), and that some
equilibrium state is achieved whereby additional water/dust deposited
in areas unfavourable for uplift can be returned to areas where uplift
can occur, e.g. basal melting of glaciers.

\begin{table}[tbhp]
\centering
\begin{tabular}{r|cc}
                                         & TL                    & nTL\\
 \hline
 Semi-major axis (AU)                    & 0.0485                & 1.00\\
 Stellar irradiance (W~m$^{-2}$)          & 881.7                 & 1361.0\\
 Stellar spectrum                        & Proxima Centauri      & The Sun\\
 Orbital period (Earth days)             & 11.186                & 365.24\\
 Rotation rate (rad~s$^{-1}$)            & 6.501$\times$10$^{-6}$ & 7.292$\times$10$^{-5}$\\
 Eccentricity                            & \multicolumn{2}{c}{0}\\
 Obliquity                               & \multicolumn{2}{c}{0}\\
 Radius (km)                             & 7160                  & 6371\\
 Gravitational acceleration (m~s$^{-1}$)  & 10.9                  & 9.81\\
\end{tabular}
\caption{{\bf Orbital and planetary properties.} Shown for the tidally-locked
  (TL) and non-tidally-locked (nTL) simulations.}
\label{tab-planparam}
\end{table}
The orbital and planetary parameters for our two template planets are
given in Table~\ref{tab-planparam}. To simplify the analysis slightly,
both planets are assumed to have zero obliquity and
eccentricity. Atmospheric parameters are given in
Table~\ref{tab-atm}. Again, for simplicity, we assume the atmospheric
composition is nitrogen dominated with trace amounts of CO$_2$ for the
control experiments investigating the role of dust on the
atmosphere. However, because of the important role of potential
biomarker gases such as oxygen, ozone and methane have in the
transmission spectrum, when discussing simulated observables we
include these gases at an abundance similar to that of present day
Earth.

\begin{table}[tbhp]
\centering
\begin{tabular}{r|cc}
 Parameter                      & Control simulations         & Synthetic observations\\
 \hline
 Mean surface pressure (Pa)     & \multicolumn{2}{c}{$10^5$}\\
 $R$ (J~kg$^{-1}$~K$^{-1}$)      & $297$                       & $287.05$ \\
 $c_p$ (J~kg$^{-1}$~K$^{-1}$)    & $1039$                      & $1005$ \\
 CO$_2$ MMR (kg~kg$^{-1}$) / ppm & $5.941\cdot10^{-4}$ / $378$ & $5.941\cdot10^{-4}$ / $391$\\
 O$_2$ MMR (kg~kg$^{-1}$) / ppm  & $0$                         & $0.2314$ / $209\cdot10^3$\\
 \multirow{2}{*}{O$_3$ MMR (kg~kg$^{-1}$) / ppm} & \multirow{2}{*}{$0$} & $2.4\cdot10^{-8}$ / $0.015$ (min) \\
                                &                             & $1.6\cdot10^{-5}$ / $9.66$ (max) \\
 CH$_4$ MMR (kg~kg$^{-1}$) / ppm & $0$                         & $1.0\cdot10^{-7}$ / $0.18$ \\
 N$_2$O MMR (kg~kg$^{-1}$) / ppm & $0$                         & $4.9\cdot10^{-7}$ / $0.32$ \\
\end{tabular}
\caption{{\bf Atmospheric parameters used in this study.} Shown for the
  baseline simulations with a simple nitrogen plus trace CO$_2$
  atmosphere, and the synthetic observations with a more
  Earth-like atmosphere. Gas quantities are given in mass 
  mixing ratio (MMR) and parts-per-million (ppm).}
  \label{tab-atm}
\end{table}

Table~\ref{tab-expts} summarises the 28 experiments and their
parameters which are described in this paper.
\begin{table}[tbhp]
\centering
\begin{tabular}{c|ccc}
 Land fraction & TL Continents      & TL Tiled                & nTL Tiled \\
 \hline
 0             & \multicolumn{2}{c}{Control}                  & Control \\
 20            & Control; Synthetic &                         & \\
 30            &                    & Control                 & Control \\
 40            & Control            &                         & \\
 60            & Control            &                         & \\
 70            &                    & Control; SI-244; SI+394 & Control; SI-161; SI+139 \\
 100           & \multicolumn{2}{c}{Control; $k_1=2$; $k_1=2$, small} & Control \\
\end{tabular}
\caption{{\bf Full list of the 28 experiments presented in this study.} The
  orbital and planetary parameters (TL, nTL) are taken from
  Table~\ref{tab-planparam}, the atmospheric parameters (Control, Synthetic) 
  are taken from Table~\ref{tab-atm} and Continents/Tiled is explained in the
  text (N.B. for $0$ and $100$\% land fraction, the Continent or Tiled
  setup is identical). Additional experiments denoted SI
  contain changes to the stellar irradiance from those quoted in
  Table~\ref{tab-planparam}, and $k_1$ vary the dust uplift as discussed
  in Supplementary Note 1. Most experiments contain two variants -- with 
  and without radiatively interactive mineral dust.}
\label{tab-expts}
\end{table}

\section*{Code availability}

The Met Office Unified Model is available for use under licence, see
\url{http://www.metoffice.gov.uk/research/modelling-systems/unified-model}.

\section*{Data availability}

All data used in this study are available from
\url{https://doi.org/10.24378/exe.2284}.

\section*{Acknowledgements}

IB and JM acknowledge the support of a Met Office Academic Partnership
secondment. We acknowledge use of the MONSooN system, a collaborative
facility supplied under the Joint Weather and Climate Research
Programme, a strategic partnership between the Met Office and the
Natural Environment Research Council. NM was part funded by a
Leverhulme Trust Research Project Grant which supported some of this
work alongside a Science and Technology Facilities Council
Consolidated Grant (ST/R000395/1). This work also benefited from the
2018 Exoplanet Summer Program in the Other Worlds Laboratory (OWL) at
the University of California, Santa Cruz, a program funded by the
Heising-Simons Foundation.

\section*{Author contributions}

IB ran the simulations and produced most of the figures and text. MJ
had the original idea and provided guidance, Figure 1 and
contributions to the text. FHL and NM provided guidance and
contributions to the text. DL investigated the role of continents as a
Masters project. JM provided scientific and technical advice. RR
produced the synthetic observations in Figure 4. KK provided technical
support.

\section*{Competing interests}

The authors declare no competing interests.

\end{document}